# Spatio-temporal measurement of ionization-induced modal index changes in gas-filled PCF by prism-assisted side-coupling

**B. M. Trabold, M. I. Suresh,\* J. R. Koehler, M. H. Frosz, F. Tani, and P. St.J. Russell**

*Max Planck Institute for the Science of Light, Staudtstrasse 2, 91058 Erlangen, Germany*
*\*Corresponding author: mallika-irene.suresh@mpl.mpg.de*

**Abstract:** We report the use of prism-assisted side-coupling to investigate the spatio-temporal dynamics of photoionization in an Ar-filled hollow-core photonic crystal fiber. By launching four different LP core modes we are able to probe temporal and spatial changes in the modal refractive index on timescales from a few hundred picoseconds to several hundred microseconds after the ionization event. We experimentally analyze the underlying gas density waves and find good agreement with quantitative and qualitative hydrodynamic predictions. Moreover, we observe periodic modulations in the MHz-range lasting for a few microseconds, indicating nanometer-scale vibrations of the fiber structure, driven by gas density waves.

## 1. Introduction

Ionization of a gas by an intense laser pulse causes complex spatio-temporal dynamics at different timescales, leading to changes in its refractive index. Over the first few nanoseconds the response is dominated by the presence of free electrons and their recombination with the ions. Hydrodynamic effects follow, lasting a few tens of nanoseconds, and finally the gas thermalizes on a µs-ms timescale. The precise timescales of these different processes depend on the experimental conditions.

Investigations of photoionization have led to a better understanding of plasmas [1,2] and filamentation [3,4], as well as the development of plasma and air waveguides [4-7] that might be exploited for laser-based free space communication in the presence of fog [8]. A good understanding of plasma dynamics is also crucial in experiments at MHz pulse repetition rates in the strong field regime, for example in high harmonic generation within femtosecond enhancement cavities [9] and the generation of broadband deep and vacuum ultraviolet via dispersive wave emission in gas-filled hollow-core photonic crystal fibers (HC-PCFs) [10].

So far, the response of a gas after ionization by fs and ns pulses has been investigated predominantly in free space at atmospheric or lower pressure, using techniques such as interferometry [4], shadowgraphy [3], electrical or sonographic probing [3,11] and x-ray absorption [2]. Although each of these techniques has its own merits and limitations, it is generally not straightforward to use any of them for studying the post-ionization dynamics of a high pressure gas confined within a small volume, such as is the case in HC-PCFs.

Recently a fiber-based interferometer was used to resolve refractive index changes along the length of a gas-filled HC-PCF and to investigate the gas dynamics after ionization triggered by a self-compressed pulse [12]. In this paper we extract additional information about the transverse spatial distribution of the photo-induced refractive index change by applying prism-assisted side-coupling [13]. This technique is highly sensitive, capable of detecting index changes as small as $10^{-6}$, and has the advantage of not relying on interferometry (which requires sophisticated stabilization techniques) but purely on phase matching, making it very robust against noise caused by air fluctuations and vibrations of mirrors in the experimental set-up.

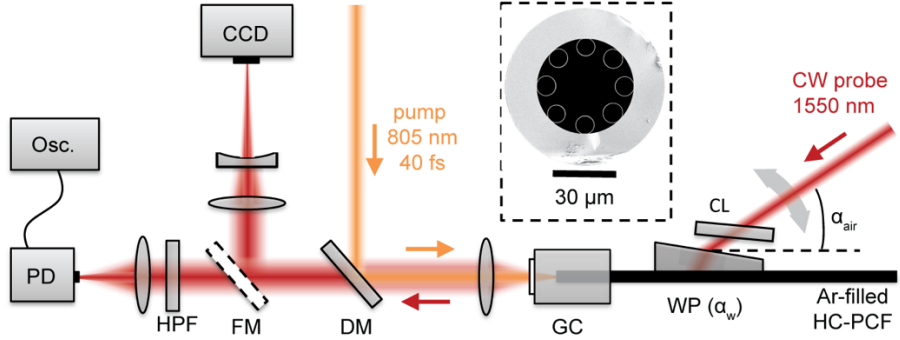

**Fig. 1:** Experimental set-up. A 20-cm-long single-ring HC-PCF (inset: scanning electron micrograph of fiber cross-section) is mounted with one end in a gas cell (GC). Pump pulses (805 nm central wavelength, 40 fs FWHM duration) are reflected off a dichroic mirror (DM) and coupled into the fiber. Continuous-wave (CW) probe light at 1550 nm is side-launched with a cylindrical lens (CL) through a wedge prism (WP; wedge angle $\alpha_w$ = 1.05°) into the fiber. A flip mirror (FM) is used to direct the emerging probe light onto a charge-coupled device (CCD) camera to image the transverse profile. When the flip mirror is absent, the probe signal can be monitored using a photodiode (PD) and an oscilloscope (Osc.).

## 2. Experimental set-up

The experimental set-up is shown in Fig. 1(a). Laser pulses of 5.4 µJ energy delivered by a Ti:sapphire laser amplifier with central wavelength 805 nm, 40 fs full-width half-maximum (FWHM) duration and 1 kHz repetition rate, are coupled into the fundamental mode of a single-ring HC-PCF (SR-PCF). A scanning electron micrograph (SEM) of the fiber cross-section is shown as an inset. The 30-µm-wide hollow core is surrounded by eight capillaries 9 µm in diameter with a wall thickness of ~190 nm, which act as anti-resonant elements. The input end of the 20 cm long fiber is placed in a gas cell filled with 7.2 bar argon, while the other end of the fiber is kept free in air at atmospheric pressure. This experimental configuration, which results in a pressure gradient, was chosen because of practical space constrains set by the shallow angle between the fiber and the probe beam. At the selected pressure, the fiber exhibits negative dispersion in the near infrared spectral region, with the zero dispersion wavelength shifting from ~550 nm at the fiber input to ~350 nm at the output. As a result, the launched pulses undergo temporal self-compression as they propagate, eventually ionizing the gas in the vicinity of the temporal focus [12,14]. According to numerical simulations (not shown) and visual inspection of the side-scattering, the pulses reach the shortest duration ~10 cm away from the fiber input, where the pressure is estimated to be ~5.5 bar [14]. This position was therefore selected to probe the evolution of the modal index after the gas is ionized.

Probe light is delivered by a continuous-wave (CW) single-frequency 1550 nm laser and coupled into the SR-PCF core using prism-assisted side-coupling [13]. A silica wedge prism (WP) with 1.05° wedge angle is mounted against the fiber, with index-matching liquid between fiber and prism so as to avoid additional refraction of the probe light. Using a rotation stage, the angle of incidence of the probe light on the prism, and thus the wavevector component $k_z$ along the fiber axis, could be controlled. A single core mode LP$_{pm}$ is then excited when $k_z$ matches its modal index $n_{pm}$. The side-coupling angle $\alpha_{air}$ at which the LP$_{pm}$ mode is excited is given by:

$$\alpha_{air} = -\alpha_w + \cos^{-1}\left[n_{gl}\cos\left(\alpha_w + \cos^{-1}(n_{pm}/n_{gl})\right)/n_{air}\right], \quad (1)$$

where $n_{gl}$ and $n_{air}$ are the refractive indices of silica and air, $\alpha_w$ is the wedge angle [13]. The core of the SR-PCF can be approximated as a simple capillary for which the effective indices of the LP$_{pm}$ modes are [14,15,16]:

$$n_{pm} = \sqrt{n_g^2 - u_{pm}^2 k_0^{-2} a_{\text{eff}}^{-2}(\lambda)},\qquad(2)$$

where $n_g$ is the refractive index of the gas, $u_{pm}$ the $m$-th zero of the $p$-th order Bessel function of the first kind, $\lambda$ the vacuum wavelength, $a_{\text{eff}}(\lambda) = a_{\text{AP}} / [1 + \lambda^2 s/(a_{\text{AP}} h)]$ the wavelength-dependent effective core radius, $a_{\text{AP}}$ the area-preserving core radius, $s = 0.08$ a fitting parameter determined by fitting experimental data of the modal indices and $h$ the thickness of the capillary walls.

In the experiment we focused probe light (power 1 W) with a cylindrical lens (20 mm focal length) to an elliptical spot of dimensions 23 μm × 1.75 mm and directed it through the prism and into the fiber core. The resulting extended focal spot made the experiment less sensitive to uncertainty as to the exact location of the temporal focus. The probe light, side-launched so as to propagate against the pump pulses, emerged into the gas cell and was transmitted through the dichroic mirror (DM) that reflected the pump pulses at 805 nm before launching them into the HC-PCF. An additional long-pass filter with 1500 nm cut-off wavelength helped attenuate residual reflections of the pump light from the uncoated window of the gas cell. The probe beam was then focused onto a fast photodiode (PD, 1.5 GHz bandwidth) and its power monitored over time using an oscilloscope (2.5 GHz bandwidth). Alternatively, a flip mirror (FM) in the beam path could be used to direct the light to an ultra-sensitive infrared charge-coupled device (CCD) camera to allow the near-field profile of the probe mode to be imaged.

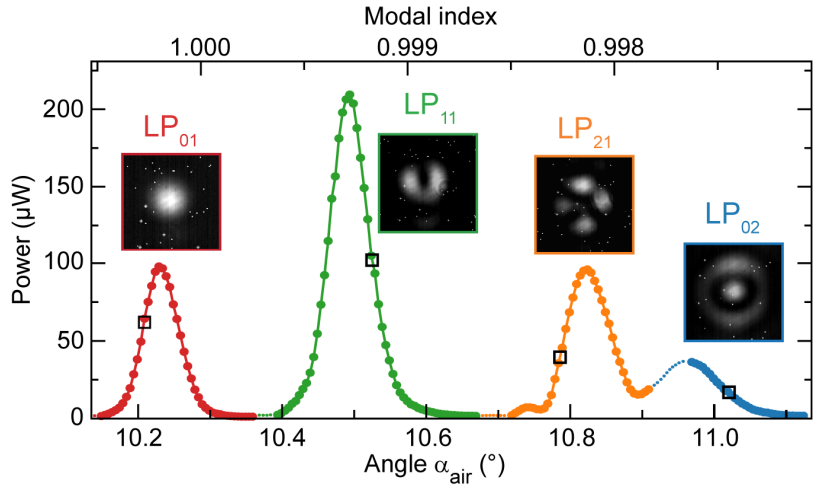

**Fig. 2:** Probe power detected at the photodiode as a function of side-coupling angle $\alpha_{\text{air}}$ (lower horizontal axis) and modal refractive index (upper horizontal axis, after Eq. (1)). The insets show the near-field profiles of the side-launched SR-PCF modes on the CCD camera. Black open squares indicate the working points for measuring plasma-related changes in the modal index. The dotted curve marks a range not used in the experiments.

To calibrate the side-launching set-up, the pump pulses were blocked and the fraction of probe light transmitted through the SR-PCF measured while varying the angle of incidence. The resulting calibration curve is shown in Fig. 2. Peaks corresponding to four distinct modes ($LP_{01}$, $LP_{11}$, $LP_{21}$ and $LP_{02}$) can be identified at different angles of incidence, the insets showing the corresponding near-field profiles recorded by the CCD camera. The mode-dependent signal power is caused by variations in modal loss and side-coupling efficiency.

### 3. Experimental results and discussion

Equation (2) indicates that any change in $n_g$ at the probing position will alter the effective refractive index and axial wavevector of the prism-coupled mode. The side-launching technique translates these changes into variations in the transmitted probe power, which can be conveniently monitored by a fast photodiode. The sensitivity of the technique is highest when, for each probed mode, the incidence angle yielding the largest slope on the calibration curve is selected as the working point (marked by open squares in Fig. 2). While the side-coupling angle $α_{air}$ is kept fixed during the measurements, the actual excitation angles of the modes and thus the calibration curve as a whole are shifted by transient changes in $n_g$.

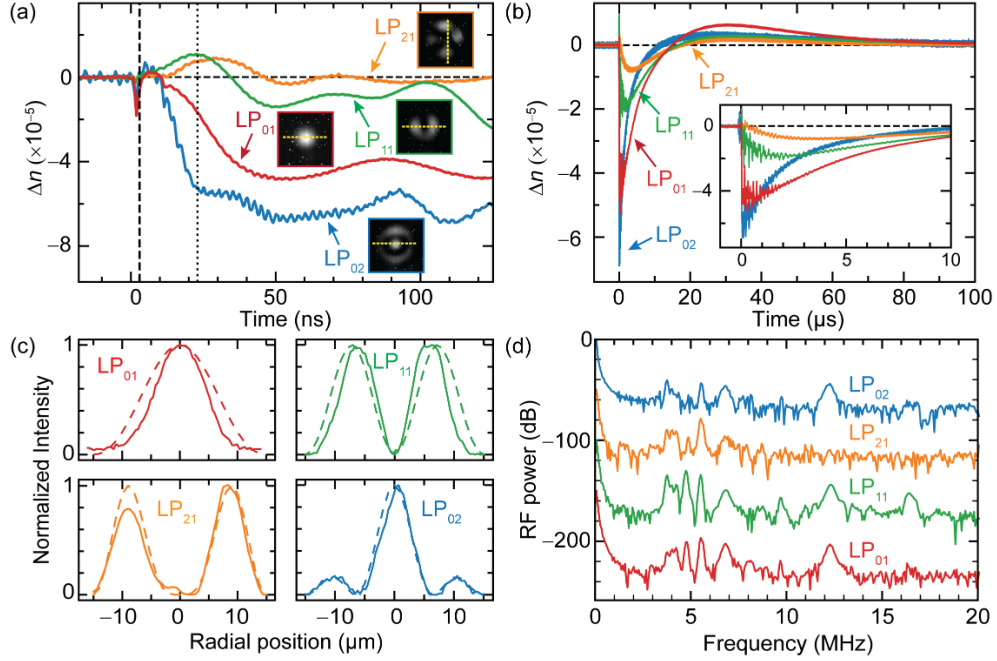

**Fig. 3:** (a) Temporal traces of the modal index change during the first 125 ns after the pump-pulse impact at $t = 0$, calculated from transmitted probe signal using the calibration curve in Fig. 2. The measured near-field mode profiles of the four excited LP modes are shown in the insets. (b) Modal index changes during the first 100 µs, the inset showing a zoom-in over the first 10 µs. (c) 1-D cuts across the modal profiles, calculated (dashed lines) and retrieved (solid lines) from the insets in (a) along the dashed yellow line. (d) Radio-frequency (RF) spectrum of the four time traces, for clarity offset from each other by 50 dB.

*3.1 Plasma-related index changes*

Figures 3(a,b) show the measured temporal response of the effective refractive index for four LP modes, calculated from the transmitted probe signal using the calibration curve in Fig. 2. To reduce noise, the traces are averaged 100 times for the $LP_{01}$, $LP_{11}$ and $LP_{21}$ modes, and 300 times for the $LP_{02}$ mode. The pump pulses arrive at the side-launching position at time $t = 0$ (temporal focus). The characteristic shape of each trace we attribute to the interplay of several effects occurring on different timescales. Initially, the refractive index of the gas filling the fiber core drops because of the presence of free electrons, approximately given by:

$$\Delta n_g \approx -N_e e^2 / (2 n_g \omega^2 \varepsilon_0 m_e), \quad (3)$$

where $N_e$ is the free electron density on the fiber axis, $e$ the electronic charge, $n_g$ the index of neutral argon, $\omega$ the angular frequency of the probe light, $\varepsilon_0$ the vacuum permittivity and $m_e$ the effective electron mass [17]. The plasma generated by the pump is localized in a narrow spot at the core center, which we estimate to be ~4 µm wide ($1/e^2$ width) using the Ammosov et al. ionization rate model [18]. As a result, the $LP_{01}$ and $LP_{02}$ modes feel a larger modal index change than the $LP_{11}$ and $LP_{21}$ modes, which carry almost no power at core center. This can be clearly seen in Fig. 3(a), where near $t = 0$ the modal index of the $LP_{01}$ mode changes by $\Delta n_{01}$ ~ $1.8 \times 10^{-5}$ over 0.5 ns (20% to 80% increase in signal), limited by the ~0.23 ns response time of the PD. The index change of the $LP_{02}$ mode is similar, while the change for the other two modes is almost an order of magnitude smaller. In order to estimate the free electron density, we can safely assume that $\Delta n_g > \Delta n_{01}$, as the width of the probe fundamental mode is much greater than the plasma spot width. By inserting $\Delta n_g > \Delta n_{01}$ into Eq. (3), we obtain $N_e > 1.7 \times 10^{16}/cm^3$. Numerical simulations usually predict $N_e$ values between $10^{17}$ and $10^{18}/cm^3$ [12]. However, the measured value is also reduced by the slow PD response and plasma-induced optical loss, which is non-negligible for light at 1550 nm.

*3.2 Pressure-wave effects*

After the gas has been ionized, gradual free electron recombination (in argon this typically takes several ns [19]) releases energy, which is mostly converted into kinetic energy. This results in local changes in gas density that evolve in time and space until the gas eventually thermalizes over tens of microseconds [3,4,6]. As the electrons recombine, a density/acoustic wave emanates from core center, with a ring-shaped front that propagates radially outwards at the speed of sound in argon (323 m/s [12]), while a slowly evolving density depression is left behind at the core center. Consequently, the effective indices of $LP_{01}$ and $LP_{02}$ drop while those of the $LP_{11}$ and $LP_{21}$ increase.

Within the first ~50 ns, the position of the density wave can be determined by correlating the traces in Fig 3(a) with the near-field images of the modes in Fig. 3(c), where for comparison we plot also the calculated modal intensity distributions [15]. After recovering from the initial plasma peak by time $t \sim 3$ ns (dashed line in Fig. 3(a)), the $LP_{02}$ effective index reaches a temporary plateau at $t \sim 23$ ns (dotted line in Fig. 3(a)), while at the same time the $LP_{11}$ index reaches its maximum. By multiplying these time intervals by the speed of sound, we obtain 6.5 µm, which coincides with the position of the first zeros of $LP_{02}$ and of the maximum of the $LP_{11}$ field intensities at 6.4 µm and 6.7 µm respectively (Fig. 3(c)). After travelling another ~7 ns, the acoustic wave reaches the position $x \sim 8.7$ µm where the $LP_{21}$ field intensity is maximum at $t \sim 30$ ns. At $t \sim 35$ ns and $x \sim 10$ µm, the acoustic wave crosses the second maximum of the $LP_{02}$ mode, with the $LP_{02}$ index beginning to decline again. Finally, at approximately $t \sim 50$ ns all modal indices reach a minimum. Theoretically, the acoustic wave has then traveled across the entire fiber core radius (15 µm) within 46 ns, which means that all modes are left in the residual density depression. After this moment, mapping the evolution of the density distribution in the fiber becomes more complex, for example, when the acoustic wave arrives at the capillaries surrounding the core, vibrational mechanical modes are impulsively excited [12]. The resulting periodic modulation of the core diameter translates into an oscillation of the modal refractive index (deducible from Eq. (2)). Part of the acoustic wave is reflected back by the capillaries, while part of it passes between them and is then reflected back by the fiber jacket. The acoustic wave bounces in this way several times back and forth within the microstructure. All these effects result in fast modulation of the modal indices over the first few microseconds, which are clearly visible in the inset of Fig. 3(b). Note however, that the rapid oscillations with a period of ~ 2.5 ns in Fig. 3(a) (particularly pronounced in the $LP_{02}$ trace) result from background electrical noise in the laboratory environment and are also present when the probe light is blocked.

On µs timescales, diffusion due to heat and density gradients causes the gas density depression in the center of the core to gradually broaden and reduce in depth. While in the first 100 ns the density depression, being localized to the core center, is only detected by the $LP_{01}$ and $LP_{02}$ probes (Fig. 3a), it slowly extends across the lobes of the $LP_{11}$ and $LP_{21}$ modes within ~5 µs. As a result, their effective indices reach a minimum significantly later than the $LP_{01}$ and $LP_{02}$ modes, namely at 2 µs ($LP_{11}$) and 3.5 µs ($LP_{21}$) – Fig. 3(b). Since the lobes of the $LP_{11}$ modes are slightly closer to the core center than those of the $LP_{21}$, the sequence of minima qualitatively visualizes the speed of thermalization. In free space, as a result of thermalization, the density depression at the center broadens monotonically on microsecond timescales until the distribution of atoms/molecules becomes spatially homogeneous [4]. However, in our experiment the gas is confined within the fiber so that, as it returns to equilibrium after ionization, an additional maximum in refractive index on a tens of microseconds timescale is observed.

### 3.3 Mechanical vibrations of capillaries

A Fourier spectrum of the time traces in Fig. 3(d) shows several peaks in the MHz range, providing further insight into the origin of the periodic index modulations. A calculation of the time taken for the generated pressure wave to be reflected from the capillaries back to the core center (a distance of 30 µm) yields a period of 93 ns, corresponding to 11 MHz. This is in reasonable agreement with the highest frequency peak (at 12.2 MHz) observed in Fig. 3(d). The peak at 6.8 MHz corresponds to a second oscillation caused by reflection of the pressure wave at the outer solid silica jacket, which travels a distance of 48 µm in 0.15 µs, yielding 6.7 MHz, in very good agreement with the measured frequency. The cluster of three narrow peaks at 3.7, 4.0, 4.2 MHz and the two separate ones at 4.8 MHz and 5.5 MHz correspond to mechanical vibrations of the capillaries and were also seen in finite element simulations (not shown). The splitting into multiple peaks we attribute to variations in wall thickness and diameter between individual capillaries.

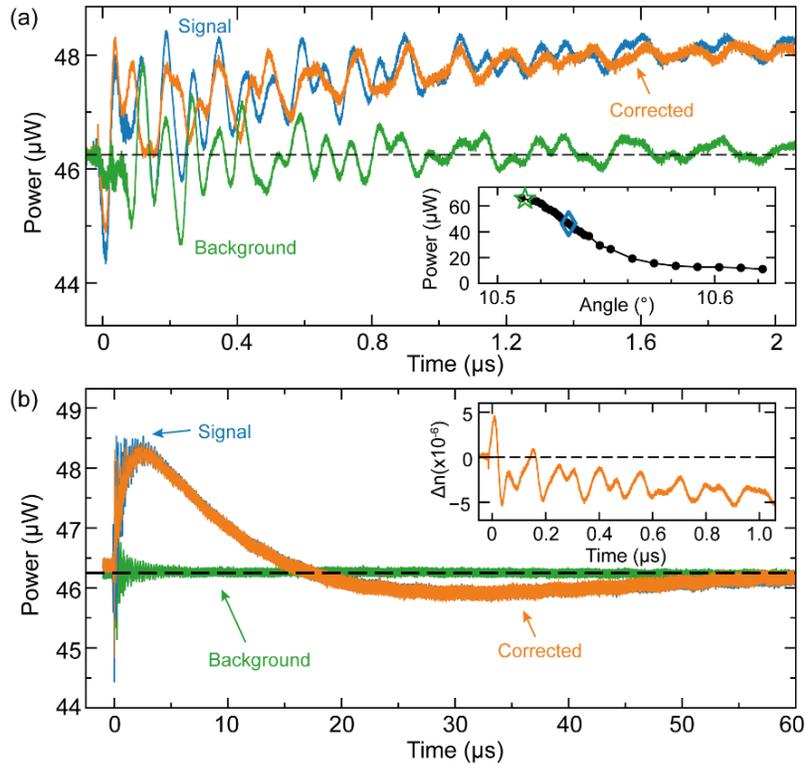

**Fig. 4:** Uncorrected signal trace (blue), scaled background trace due to coupling efficiency changes (green) and background-corrected trace (orange) only including effects of modal index changes (a) in the first 2 µs and (b) over 60 µs, measured for $LP_{11}$ mode. Inset of (a): Working points for measuring uncorrected signal trace (blue diamond) and for background trace (green star) marked on the calibration curve. Inset of (b): Calculated modal index change from the background-corrected curve.

### 3.4 Corrections for variations in coupling efficiency

We now consider in more detail the effects of the mechanical oscillations of the capillaries and the density wave resonating within the fiber core. Although both of them modulate the modal index as mentioned above, they also modify the refraction and reflection of the side-launched probe beam, and can thus alter the side-coupling efficiency, reducing the accuracy of the measured index changes. The contribution of these effects can be eliminated by measuring a background trace with the working point set to the top of the calibration curve, where the slope is flat and any index change only minimally affects the transmitted probe power. At the same time any variations in transmitted power caused by changes in coupling efficiency are maximized.

We now explore this effect for the $LP_{11}$ mode, for which the largest measured modal index change is $\Delta n \approx -2.6\times10^{-5}$ (Fig. 3), corresponding to a change in angle $\Delta\alpha = 0.0075°$, which exceeds the region within which the slope in the calibration curve is approximately zero. In order to examine the pure effect of changes in side-launching efficiency, therefore, the pump pulse energy was reduced from 5.4 to 5 µJ, resulting in a five times reduction in modal index and $\Delta\alpha = 0.0015°$. At the background working point (green star in the inset of Fig. 4(a)), the coupled power is $P_{BWP} = 66$ µW, while at the signal working point (blue diamond), it is only $P_{SWP} = 46$ µW. Consequently, we scale the measured background curve $P_{back}(t)$ down by a factor of $P_{SWP}/P_{BWP} \approx 0.70$ (green trace in Fig. 4) before subtracting it from the measured signal trace $P_{sig}(t)$ (blue trace). This results in $P_{corr}(t) = P_{sig}(t) - (0.70\, P_{back}(t) - P_{SWP})$ for the corrected signal (orange curve in Fig. 4), including only variations due to modal index changes.

As expected, the background oscillates over the first few microseconds, due to the density wave resonating in the fiber core and by capillary vibrations, but is zero apart from that. This means that the temporal traces in Fig. 3, for which no background was subtracted, are only slightly affected by the periodic modulation background over the first few microseconds. To obtain quantitative information on these vibrations, we converted the background-corrected signal into a modal index change (inset of Fig. 4(b)). Using the corrected data together with Eq. (2), we could estimate the deflection of the capillaries. Filtering out in Fourier space the peak corresponding to capillary vibrations and transforming it back into the time domain, we obtain $-0.35\times10^{-6}$ as the maximum modulation of the modal index. Inserting this value into Eq. (2), we obtain a core diameter change of 3 nm, corresponding to a capillary wall deflection of 1.5 nm. Using the $LP_{21}$ mode as probe, the measurements yield a much larger deflection (12 nm), which we attribute to the fact that the lobes of the $LP_{21}$ penetrate deeper between the surrounding capillaries. As a result, the assumption that the index change is due only to a change in core diameter is no longer valid. Note however, that this measurement was made at reduced pump energy; the capillary wall deflection will be much larger at higher pump pulse energies.

## 4. Conclusions

The prism-assisted side-coupling technique allows monitoring, over timescales from a few ns to tens of µs, of the dynamics following fs ionization in gas-filled hollow-core photonic crystal fiber. Making use of the different transverse profiles of four different LP core modes, the spatial positions of the pressure wave generated after the free electron recombination can be extracted. Since the gas is confined at a high pressure within a small hollow fiber core, the response is more complex than in free-space studies, because it is affected also by gas density wave vibrations and mechanical vibrations of the capillary walls.